\documentclass[aps, showpacs, showkeys,nofootinbib,floatfix]{revtex4}

\usepackage{amssymb}
\usepackage{amsmath}
\usepackage{graphicx}
\usepackage{hyperref}

\begin{document}

\title{Real Scalar Field Scattering with Polynomial
Approximation around Schwarzschild-de Sitter Black-hole}

\author{Molin Liu}
\email{mlliudl@student.dlut.edu.cn}
\author{Hongya Liu}
\email{hyliu@dlut.edu.cn}
\author{Jingfei Zhang}
\author{Fei Yu}

\affiliation{School of Physics and Optoelectronic Technology, Dalian
University of Technology, Dalian, 116024, P. R. China}

\begin{abstract}
As one of the fitting methods, the polynomial approximation is
effective to process sophisticated problem. In this paper, we employ
this approach to handle the scattering of scalar field around the
Schwarzschild-de Sitter black-hole. The complex relationship between
tortoise coordinate and radial coordinate is replaced by the
approximate polynomial. The Schr$\ddot{o}$dinger-like equation, the
real boundary conditions and the polynomial approximation construct
a full Sturm-Liouville type problem. Then this boundary value
problem can be solved numerically according to two limiting cases:
the first one is the Nariai black-hole whose horizons are close to
each other, the second one is when the horizons are widely
separated. Compared with previous results (Brevik and Tian), the
field near the event horizon and cosmological horizon can have a
better description.
\end{abstract}



\maketitle

\section{Introduction}
In the black-hole physics Hawking radiation\cite{Hawking1} is
always a very important conception which indicates that
black-holes are not perfect black, but radiate thermally and
eventually explode. An in-depth discussion of derivation can be
obtained in Ref.\cite{Brout} As for the recent research, one can
refer to Ref.\cite{CP} Many researchers have developed various
methods and techniques to study the black-hole by using the
radiating particles, such as the simple Klein-Gordon particles and
Dirac particles (for some early works, see Damour and
Ruffini\cite{Damour} and Chandrasekhar\cite{Chandrasekhar}
respectively). Recently, as for scalar field, Higuchi et
al.\cite{Higuchi} and Grispino et al.\cite{Grispino} gave its
solution outside a Schwarzschild black-hole, Brady et
al.\cite{Brady} studied the Schwarzschild-de Sitter case and Guo
et al.\cite{Guo} made further studies in the
Reissner-Nordstr$\ddot{o}$m-de Sitter one.

The Schwarzschild-de Sitter (SdS) space is a spherically symmetric
system.\cite{Rindler} It can be treated as a small Schwarzschild
black-hole embedded in de Sitter universe. In this space there are
two horizons: one is inner black-hole horizon $r_{e}$ and the
other is outer cosmological horizons $r_{c}$. In 2001, Brevik and
Simonsen\cite{Brevik} gave a massless scalar field solution by
tangent approximation which contains an explicit tangent function.
Viewing from the global frame, this method matches $r$ with
$\tilde{r}$ very well, where $r$ is the radial coordinate and
$\tilde{r}$ is a fitting function. However, its insufficiency is
the weak fitting near the two horizons. Even in the intermediate
zone, the fitting $r$ with $\tilde{r}$ is not precise enough. In
their paper,\cite{Brevik} they have studied for two extreme cases
whose horizons are either very close to each other or lie very far
away. Afterwards, Tian et al\cite{Tian} used a polynomial
approximation containing 20 monomials and gave another different
numerical solution only in the extreme Nariai black-hole. This
useful polynomial approximation is more precise than the tangent
approximation. Especially, in the leading intermediate zone, the
fitting $r$ with a polynomial makes a good match. However, this
type of approximation rapidly deteriorates near the horizons, i.e.
inappropriate boundary conditions are used. On the other hand, the
widely separated horizons case is not considered in
Ref.\cite{Tian} Considering the above situation, we re-study the
scattering of a scalar field with polynomial approximation.

This paper is organized as follows: in Section 2, we present the
Schwarzschild-de Sitter space and point out the positions of
black-hole horizon and cosmological horizon. In Section 3, by the
polynomial approximation, the full Sturm-Liouville type problems
are solved for two extreme cases. Section 4 is a conclusion. We
adopt the signature $(+, -, -, -)$ and put $\hbar$, $c$ ,and $G$
equal to unity. The same setting of field parameters is employed
in Ref.\cite{Brevik}
\section{Scalar Field in Schwarzschild-de Sitter Space}
The spherically symmetric metric of the Schwarzschild-de Sitter
space\cite{Rindler} is given by
\begin{equation}\label{metric}
    d s^2 = f(r) d t^2 - \frac{1}{f(r)} d r^2 - r^2 \left(d
    \theta^2 +\sin^2\theta d \phi^2\right),
\end{equation}
where
\begin{equation}
f(r)=1-\frac{2M}{r}-\frac{\Lambda}{3}r^2,\label{f-function}
\end{equation}
with the black-hole mass $M$ and the cosmological constant
$\Lambda$. It is an exact exterior solution of the Einstein field
equations for a spherical mass distribution
\begin{equation}\label{EinEQ}
    R_{\mu\nu} - \frac{1}{2} g_{\mu\nu} R + \Lambda g_{\mu\nu} = 8
    \pi G T_{\mu\nu}.
\end{equation}
Here, we take the cosmological constant $\Lambda$ as a free
parameter. The similar process can be found in
Refs.\cite{Brevik,Tian,Liu00} This space is bounded by two
horizons --- an inner horizon (black-hole horizon) and an outer
horizon (cosmological horizon). Under the limit $\Lambda
\longrightarrow 0$, this metric has exactly the same line-element
as the Schwarzschild space. But in the limit of $M \longrightarrow
0$, it reduces to the de Sitter one.

Mathematically, expression (\ref{f-function}) can be rewritten as
\begin{equation}
f(r)=\frac{\Lambda}{3r}(r-r_{e})(r_{c}-r)(r-r_{o}). \label{re-f function}%
\end{equation}
The singularity of metric (\ref{metric}) is determined by $f(r) =
0$. The solutions to this equation are these to inner horizon
$r_{e}$ and outer horizon $r_{c}$, as well as a negative solution
$r_{o}=-(r_{e}+r_{c})$. The last one has no physical meaning. Here
we only consider the positive solutions. The positions of $r_{c}$
and $r_{e}$ are given by
\begin{equation}
\left\{
\begin{array}{c}
r_{c} = \frac{2}{\sqrt{\Lambda}}\cos\eta ,\\
r_{e} = \frac{2}{\sqrt{\Lambda}}\cos(120^\circ-\eta),\\
\end{array}
\right.\label{re-rc}
\end{equation}
where $\eta = 1/3 \arccos(-3M\sqrt{\Lambda})$ with $30^\circ
\leq\eta\leq 60^\circ$. The real physical solutions are accepted
only if $\Lambda$ satisfies $\Lambda M^2\leq 1/9$.\cite{Liu1} If
the cosmological constant $\Lambda$ reaches its maximum, the
Nariai black-hole appears.\cite{Nariai}

A massless scalar field $\Phi (t, r, \theta, \phi)$ is considered
here. Using the separable solutions\cite{Jensen}
\begin{equation}
\Phi=\frac{1}{\sqrt{4\pi\omega}}\frac{1}{r}R_{\omega}(r,t)Y_{lm}(\theta,\phi),\label{wave
function}
\end{equation}
the scalar field equation
\begin{equation}
\square\Phi=0,\label{Klein-Gorden equation}
\end{equation}
is decomposed into two differential equations:
\begin{eqnarray}
\nonumber  -\frac{1}{f(r)} r^2\frac{\partial^2}{\partial
t^2}\left(\frac{R_{\omega}}{r}\right)&+&\frac{\partial}{\partial
r}\left(r^2
 f(r)\frac{\partial}{\partial{r}}\left(\frac{R_{\omega}}{r}\right)\right) \\
  &-& l(l+1)\frac{R_{\omega}}{r}=0,\label{radius-t-equation}
\end{eqnarray}
\begin{equation}\label{sph.equ.}
\frac{1}{\sin\theta}\frac{\partial}{\partial\theta}\left(\sin\theta\frac{\partial{Y_{lm}}}{\partial\theta}\right)+\frac{1}{\sin^2\theta}\frac{\partial^2
Y_{lm}}{\partial{\phi^2}}=-l(l+1)Y_{lm},
\end{equation}
where $R_{\omega}(r, t)$ is the time-dependent radial function and
$Y_{l \omega} (\theta, \phi)$ is the spherical harmonics function.
Equation (\ref{radius-t-equation}) determines the evolution of
evaporating black-hole. It is necessary to eliminate the time
variable by the Fourier component $e^{-i \omega t}$ via
 \begin{equation}
R_{\omega}(r,t)\rightarrow\Psi_{\omega l}(r) e^{-i\omega t}.
 \end{equation}
So Eq.(\ref{radius-t-equation}) can be rewritten as%
 \begin{equation}
 \left[-f(r)\frac{d}{dr}(f(r)\frac{d}{dr})+V(r)\right]\Psi_{\omega
 l}(r)=\omega^2\Psi_{\omega l}(r),\label{radius equ. about r}
 \end{equation}
whose potential function is given by%
\begin{equation}
V(r)=f(r)\left[\frac{1}{r}\frac{df(r)}{dr}+\frac{l(l+1)}{r^2}\right].\label{potential-of-r}
\end{equation}

Now we introduce the tortoise coordinate%
\begin{equation}
x=\frac{1}{2M}\int\frac{dr}{f(r)}.\label{tortoise }
\end{equation}
The tortoise coordinate can be expressed by surface gravity as
follows:
\begin{eqnarray}\label{tor-grav-sf}
 \nonumber x &=& \frac{1}{2M}\bigg{[}\frac{1}{2K_{e}}\ln\left(\frac{r}{r_{e}}-1\right)-\frac{1}{2K_{c}}\ln\left(1-\frac{r}{r_{c}}\right) \\
  &+& \frac{1}{2K_{o}}\ln\left(1-\frac{r}{r_{o}}\right)\bigg{]},
\end{eqnarray}
where%
\begin{equation}
K_{i}=\frac{1}{2}\left|\frac{df}{dr}\right|_{r=r_i}.
\end{equation}
Explicitly, we have%
\begin{eqnarray}
  K_{e}=\frac{(r_{c}-r_{e})(r_{e}-r_{o})}{6r_{e}}\Lambda, \\
  K_{c}= \frac{(r_{c}-r_{e})(r_{c}-r_{o})}{6r_{c}}\Lambda,\\
  K_{o}= \frac{(r_{o}-r_{e})(r_{c}-r_{o})}{6r_{o}}\Lambda.
\end{eqnarray}
By tortoise coordinate transformation (\ref{tortoise }), the radial equation (\ref{radius equ. about r}) can be written in the $''\text{Regge - Wheeler}''$ form%
\begin{equation}
\left[-\frac{d^2}{dx^2}+4M^2V(r)\right]\Psi_{\omega
l}(x)=4M^2\omega^2\Psi_{\omega l}(x),\label{radius-equation}
\end{equation}
which has the form of Schr$\ddot{o}$dinger equation of quantum
mechanics. It is usually called Schr$\ddot{o}$dinger-like
equation. The incoming or outgoing particle flow between inner
horizon $r_{e}$ and outer horizon $r_{c}$ is reflected and
transmitted by the potential barrier $V(r)$. The evolution of wave
solution $\Psi_{\omega l}$ of massless scalar field is also
determined by potential $V(r)$.
\section{Full Sturm-Liouville Type Problem}
\begin{figure}
\includegraphics[width=3 in]{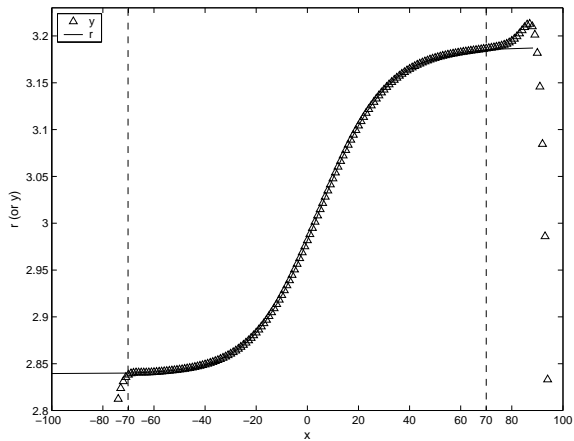}\\
\caption{The radial coordinate r (full line) and polynomial
approximation $y$ (triangle-up) versus the tortoise coordinate $x$
with $\Lambda = 0.11$.}
\end{figure}
According to Eqs. (\ref{re-f function}) and
(\ref{potential-of-r}), the potential $V(r)$ disappears near
black-hole horizon $r_{e}$ and cosmological horizon $r_{c}$. So
the potentials near horizons are given as follows:
\begin{equation}\label{Boundary-potential}
    V(r_{e})=V(r_{c})=0.
\end{equation}
Hence near the horizons, Eq.(\ref{radius-equation}) reduces to
\begin{equation}\label{Boundary-Equ}
    \left[\frac{d^2}{dx^2}+4M^2\omega^2\right]\Psi_{\omega
    l}(x)=0.
\end{equation}
Absolutely, its solutions are $e^{\pm i2M\omega x}$ or their
comprehensive form. Taking the real scalar field\cite{Brevik,Tian}
into account, we choose the real part of its solutions as the
boundary condition
\begin{equation}\label{Boundary-conditions}
   \Psi_{\omega l}=\cos(2M\omega x).
\end{equation}

\begin{figure}\label{f2}
\includegraphics[width=3 in]{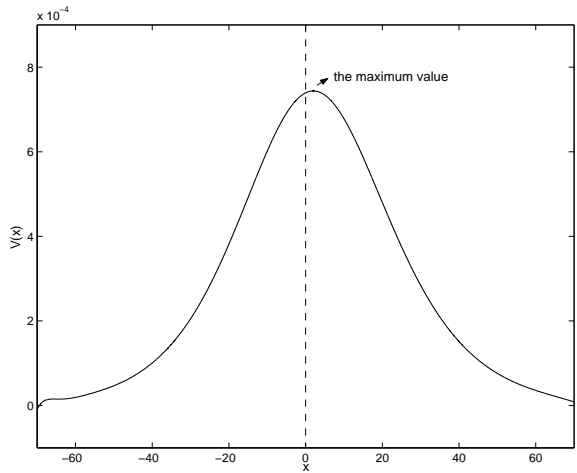}\\
\caption{The potential $V(x)$ versus tortoise coordinate $x$ with
$\Lambda = 0.11$, $M = 1$, $l = 1$, where the maximum value is
represented by the black point.}
\end{figure}
There are two coordinates --- radial coordinate $r$ and tortoise
coordinate $x$ contained in Eq.(\ref{radius-equation}). However,
the source transformation expression (\ref{tor-grav-sf}) is too
complicated to invert it to the form of $r = r (x)$. In order to
solve Eq.(\ref{radius-equation}) conveniently, it is necessary to
use an approximate method for transition
\begin{equation}\label{1}
    r \approx \tilde{r}(x),
\end{equation}
where the radial coordinate $r$ is replaced by the fitting
function $\tilde{r}$ containing the tortoise coordinate $x$. Here
we use the polynomial approximation\cite{Tian} to fit $r$ with
$\tilde{r}$. This method involves complicated polynomials unlike
the explicit tangent approximation.\cite{Brevik} For any given
value of the parameter $\Lambda$, we can always find an
appropriate approximate method from the above by adjusting the
parameters. The difficulty is how to obtain the fitting function
(\ref{1}). Here, we consider the polynomial
approximation\cite{Tian}
\begin{equation}\label{polynomal}
y = \tilde{r} = \sum ^{N}_{i = 0}  a_{i} x^{i},
\end{equation}
where ${a_{i}}$ is the coefficient and $N$ is the degree of
polynomial. In the various fitting functions, the greatest
advantage of the polynomial approximation is that it can obtain
the optimal approximation by the adjustable parameter $N$. One
should note that it is wrong to use the bigger $N$ to gain a more
accuracy approximation. $N$ must be chosen according to the
fitting interval. By combining the potential $V (x)$, the
Schr\"{o}dinger-like equation (\ref{radius-equation}), fitting
function (\ref{polynomal}) with the boundary conditions
(\ref{nobc}), we can present a full Sturm-Liouville type problem.
This kind of boundary value problem is usually used to solve the
field equation, such as Refs.\cite{Brevik, Tian, Liu00} Note that
the same setting of field parameters as that in Ref.\cite{Brevik}
are adopted in the subsections.
\subsection{More Exact Boundary Conditions for Nariai Case: $\Lambda = 0.11$}
\begin{figure}
\includegraphics[width=3 in]{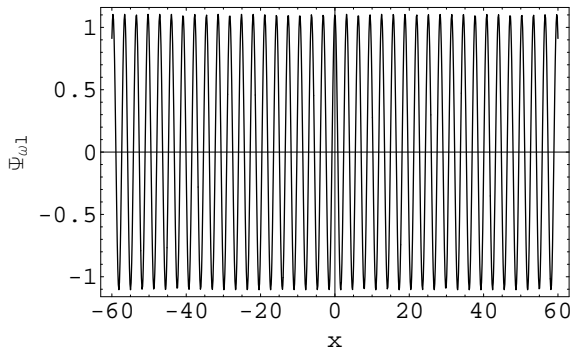}\\
\caption{The wave function $\psi_{\omega l}$ versus the tortoise
coordinate $x$ with $\Lambda = 10^{-3}$, $M =1$ and $l = 1$.}
\end{figure}
Nariai solution has been discovered by Nariai,\cite{Nariai} which
is the exact solution to the Einstein equation with $\Lambda
> 0$ without a Maxwell field. Its topological structure is a
(1+1)-dimensional dS spacetime with a round 2-sphere of fixed
radius, i.e., $d S_2\times S^2$. We adopt the same value
$\Lambda=0.11$ as appeared in Refs.\cite{Brevik,Tian} Then
substituting $\Lambda$ into Eq.(\ref{re-rc}), we find the inner
horizon is $r_{e}= 2.8391M$ and the outer horizon is $r_{c} =
3.1878M$. Just as in Ref.\cite{Tian}, we employ the same
polynomial approximation here. The coefficients $\{a_{i}\}$ are
showed in Table 1.
\begin{table*}\label{table}
\begin{center}
\caption{The coefficients of degree 20 polynomial with $\Lambda =
0.11$.}
\begin{tabular}{|l|l|l|}
     \hline\hline
     $a_{0} = 2.9817$ &$a_{1} = 6.5107 \times 10^{-3}$&$a_{2} = 4.0912 \times 10^{-5}$\\
     \hline
     $a_{3} = -2.9913 \times 10^{-6}$&$a_{4} = -3.4895 \times 10^{-8}$&$a_{5} = 1.6009 \times 10^{-9}$\\
     \hline
     $a_{6} = 2.3413 \times 10^{-11}$&$a_{7} = -8.0083\times 10^{-13}$&$a_{8} = -1.1964 \times 10^{-14}$\\
     \hline
     $a_{9} = 3.3845 \times 10^{-16}$&$a_{10} = 4.3110 \times 10^{-18}$&$a_{11} = -1.0899 \times 10^{-19}$\\
     \hline
    $a_{12} = -1.0120 \times 10^{-21}$&$a_{13} = 2.4364 \times 10^{-23}$&$a_{14} = 1.4031 \times 10^{-25}$\\
    \hline
    $a_{15} = -3.4329 \times 10^{-27}$&$a_{16} = -9.5242 \times 10^{-30}$&$a_{17} = 2.6439 \times 10^{-31}$\\
    \hline
    $a_{18} = 1.5652 \times 10^{-34}$&$a_{19} = -8.0402 \times 10^{-36}$&$a_{20} = 4.3262 \times 10^{-39}$\\
     \hline
     \hline
\end{tabular}
\end{center}
\end{table*}

The boundary conditions in Ref.\cite{Tian} came directly from the
original work in Ref.\cite{Brevik},
\begin{equation}\label{obc}
\Psi_{\omega l}(x)|_{x=-100} = \Psi_{\omega l}(x)|_{x=100} =
\cos(200 M \omega).
\end{equation}
However, it is not appropriate to use Eq.(\ref{obc}) directly as
the boundary conditions do not consider the new approximation. The
intervals of the boundary conditions should be in accord with the
fitting intervals. Now, we present the inappropriate boundary
conditions in previous polynomial approximation.\cite{Tian} One
can treat the polynomial (\ref{polynomal}) as a function varying
with $x$ for $N = 20$. The coefficients are listed in Table.1. The
functional images of Eqs.(\ref{polynomal}) and (\ref{tor-grav-sf})
both are drawn in Fig.1. It is shown that $\tilde{r}$ (or $y$) and
$r$ match exactly in the intermediate zone. However, near the
horizons $r_{e}$ and $r_{c}$, the polynomial behavior takes over
and the approximation quickly deteriorates.
\begin{figure}
\includegraphics[width=3 in]{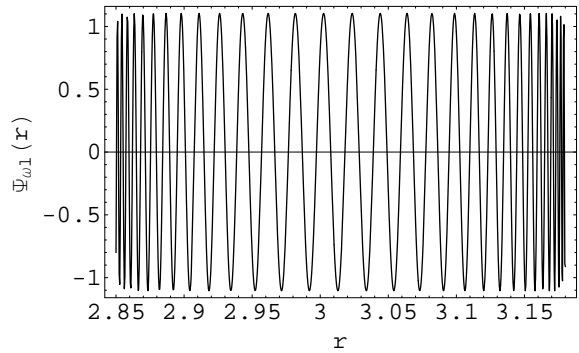}\\
\caption{The variation of the wave function $\psi_{\omega l}$ with
the radial coordinate $r$ for $\Lambda = 10^{-3}$, $M =1$ and $l =
1$.}
\end{figure}
Especially, there are significant differences between $r$ and $y$
in the two intervals, $[-70,-100]$ and $[70,100]$, along the
horizontal axis. Because of the unnecessary intervals, the
singular peak has arisen in the waves (one can refer to Figs.4 and
5 in Ref.\cite{Tian}). This problem can be solved by reducing the
interval $x$ from $[-100,100]$ to $[-70,70]$. Hence, the effective
interval in the radial direction has been changed from $[2.8391M,
3,1878M]$ to $[2.8382M, 3.1865M]$. After removing the useless
intervals: $[-70,-100]$ and $[70,100]$, we obtain another exact
boundary conditions
\begin{equation}\label{nobc}
\Psi_{\omega l}(x)|_{x=-70} = \Psi_{\omega l}(x)|_{x=70} =
\cos(140 M \omega).
\end{equation}
The potential $V(x)$ of Nariai black-hole is plotted in the range
of $-70\leq x\leq 70$ in Fig.2, with maximum value $V = 7.4381
\times 10^{-4}$ corresponding to $x = 2.0352$. By using
Mathematica software in book,\cite{Wolfram} one can solve it
numerically as a boundary value problem, where the command NDSolve
is used. The amplitude versus the tortoise coordinate is shown in
Fig.3. It is seen that the solution $\Psi_{\omega l}(x)$ is
similar to a harmonic wave without considering the decay factor
$1/r$ in the ansatz (\ref{wave function}). The radial equation
(\ref{radius-t-equation}) is transformed into a standard wave
equation by useful tortoise transformation (\ref{tortoise }). With
the real boundary conditions (cosine functions), the harmonic wave
arises naturally. Taking into account the actual case, we also
plot the amplitude versus $r$ in Fig.4. This diagram illustrates
clearly that waves stack up near $r_{e}$ and $r_{c}$.
\begin{figure}
\includegraphics[width=3 in]{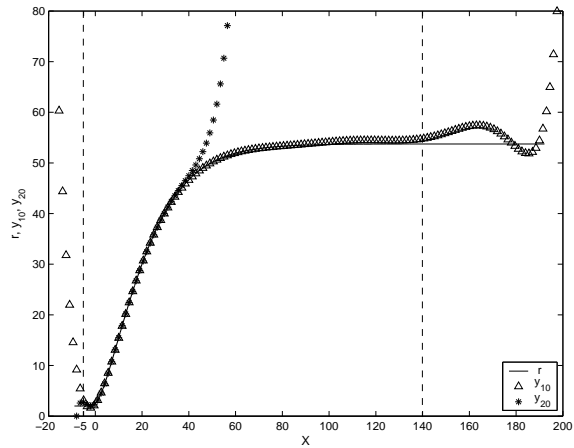}\\
\caption{The radial coordinate $r$ (full line), the 10-th polynomial
$y_{10}$ (triangle-up) and the 20-th polynomial $y_{20}$ (star)
versus the tortoise coordinate $x$ with $\Lambda =
10^{-3}$.}\label{ryx_0.001}
\end{figure}

For the case of Brevik's work, because the tangent approximation
does not work well near the two horizons (see Fig.3 in
Ref.\cite{Brevik}), waves do not pile up near the outer horizon
$r_{c}$ (see Fig.6 in Ref.\cite{Brevik}). For the case of Tian's
work, there is a singular peak in waves near $x = 2.8582$ (or $r
\sim 3$) (see Figs.4 and 5 in Ref.\cite{Tian}). The occurrence of
this singular peak is due to inappropriate boundary conditions
(\ref{obc}) chosen. In this paper, the precise polynomial
approximation is kept, but the boundary conditions (\ref{obc}) are
replaced by the new exact ones (\ref{nobc}). However, viewing from
the numerical solutions shown in Figs.3 and 4, we can say that the
afore-mentioned deficiencies have been remedied.
\begin{figure}
\includegraphics[width=3 in]{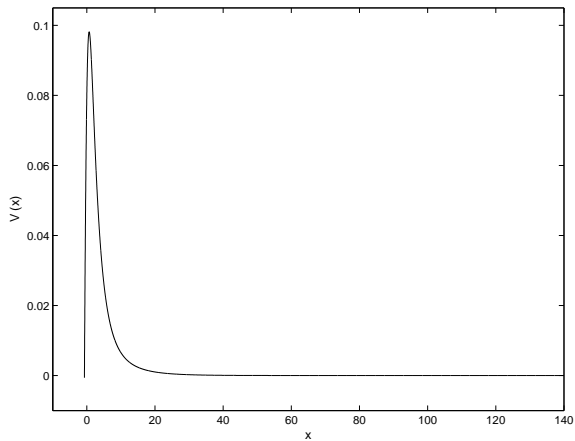}\\
\caption{The potential $V(x)$ versus tortoise coordinate $x$ with
$\Lambda = 10^{-3}$, $M=1$,  $l=1$.}\label{potential_0.001}
\end{figure}

\subsection{Widely Separated Horizons Case: $\Lambda = 0.001$}
The accelerating universe phenomenon can be easily explained by a
repulsive force produced by non-zero and positive cosmological
constant $\Lambda_{0}\sim 10^{-52} m^{-2}$.\cite{Schidt} The
cosmological constant has very interesting gravitational effects
on various astrophysical scales such as the gravitational lensing
statistics of extragalactic surveys,\cite{Quast} large-scale
velocity flows,\cite{Zehavi} the effects on observation in small
system (Galactic,\cite{Whitehouse} Planetary,\cite{Cardona} and
Solar\cite{Kagramanova}). Considering the widely separated
horizons case, we take the cosmological constant $\Lambda =
10^{-3}$, which is employed as a general setting in many
works.\cite{Brevik,Liu00} Of course, the other value of
cosmological constant subjected to the condition $\Lambda M^2 \leq
1/9$ also can be adopted in principle. Here we take the same value
as that in Ref.\cite{Brevik}

In the polynomial approximation, the number of terms can be
selected at random in principle. Two points determine one line or
an approximating polynomial of degree 1; three points determine an
approximating polynomial of degree 2, and so on; while $n + 1$
points determine the approximating polynomial of degree n.
However, the polynomial with higher degree presents some defective
numerical characteristics. With increasing degree, the fitting
curve becomes lack of smoothness because the higher degree
polynomial can be differentiated many times before it reduces to
zero. We thus choose a right polynomial rather than a higher
degree one.
\begin{figure}
\includegraphics[width=3 in]{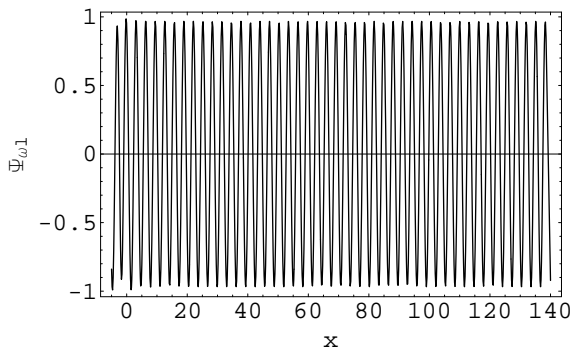}\\
\caption{The variation of the wave function $\psi_{\omega l}$ with
the tortoise coordinate $x$ for $\Lambda = 10^{-3}$, $M =1$ and $l =
1$.}\label{wave_x_0.001}
\end{figure}

The polynomials with 20-th degree ($y_{20}$) and 10-th degree
($y_{10}$) are plotted in Fig.5. We can read that the fitting
interval of $y_{10}$ is $[-5, 140]$ and the fitting interval of
$y_{20}$ is $[-5, 40]$. Obviously, the former is much wider than
the latter. Here, we adopt the 10-th degree polynomial as fitting
function (\ref{polynomal}). The coefficients are listed in Table
2. Using this approximation, the potential $V(x)$ is plotted in
Fig.6. The curve peak approaches the event horizon $r_{e}$ and is
far apart from the cosmological horizon $r_{c}$. Like the former
case, this kind of boundary value problem can be solved
numerically\cite{Wolfram} with a more general setting which is
more faithfully representing our world. The numerical solutions
$\Psi_{\omega l}(x)$ and $\Psi_{\omega l}(r)$ are shown in Figs.7
and 8, respectively. Obviously, with decreasing $\Lambda$ the
waves become much sparser near $r_{e}$ and much denser near
$r_{c}$. Otherwise, since all parts of the potential are in the
region of $x > 0$, the wave solutions $\Psi (x)$ and $\Psi (r)$
are concentrated in the positive horizontal axis too.
\begin{table*}
\begin{center}
\caption{The coefficients in the approximating polynomial of degree
10}
\begin{tabular}{|l|l|l|}
     \hline\hline
     $a_{0} = 2.3895$ &$a_{1} = 0.64096$&$a_{2} = 0.11802$\\
     \hline
     $a_{3} = -6.7621 \times 10^{-3}$&$a_{4} = 1.8818 \times 10^{-4}$&$a_{5} = -3.2408 \times 10^{-6}$\\
     \hline
     $a_{6} = 3.6365 \times 10^{-8}$&$a_{7} = -2.6487\times 10^{-10}$&$a_{8} = 1.2019 \times 10^{-12}$\\
     \hline
     $a_{9} = -3.0754 \times 10^{-15}$&$a_{10} = 3.3792 \times 10^{-18}$&\\
     \hline
     \hline
\end{tabular}
\end{center}
\end{table*}
\begin{figure}
\includegraphics[width=3 in]{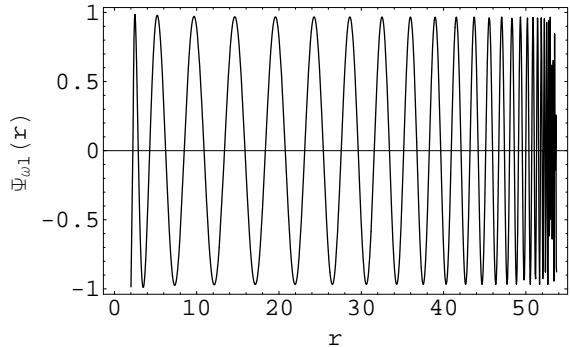}\\
\caption{The variation of the wave function $\psi_{\omega l}$ with
the radial coordinate $r$ for $\Lambda = 10^{-3}$, $M =1$ and $l =
1$.}\label{wave_r_0.001}
\end{figure}
\section{Conclusion}
We have solved the real scalar field numerically with the
polynomial approximation. Unlike the previous original
work,\cite{Tian} we have not only surveyed a more general setting
of field parameters, i.e. the widely separated horizons case, but
also selected more precise boundary conditions. We summarize what
have been achieved as follows.

1 Polynomial approximation is an important and comprehensible
approximate method. In this paper this method has been used to fit
the radial coordinate with the tortoise coordinate. Unlike the
tangent approximation,\cite{Brevik} one main merit of this method
is that there is an adjustable parameter, the degree of the
polynomial. The degree of the polynomial must be selected to
ensure that we can obtain the maximum fitting interval. From the
analysis in this paper, we find that the degree of the polynomial
should be reduced with increasing cosmological constant. When we
consider two extreme cases with $\Lambda = 0.11$ and $\Lambda =
0.001$, the degree should be reduced from 20 to 10.

2 As for the Nariai black-hole, because the potential barrier $V(r)$
(\ref{potential-of-r}) vanishes near the two horizons,
Eq.(\ref{Boundary-conditions}) becomes a universal boundary
condition for real solution case. But the different fitting
intervals need different boundary conditions. So it has to choose a
well and suitable one according to the quality of fitting. In this
paper, we have chosen another exact boundary condition. Although the
previous polynomial approximation is more accurate to match $r$ with
$\tilde{r}$ in the intermediate zone, the interval of $x$ does not
keep the original one ($[-100, 100]$). It is illustrated clearly in
Fig.1. Obviously, when the tortoise coordinate $x$ is in the range
of $[-100,-70]$ or $[70,100]$, the approximation (\ref{polynomal})
quickly deteriorates. So it is necessary to use the new interval
$[-70, 70]$ to replace the previous one $[-100, 100]$. As mentioned
in the above sections, by tortoise coordinate transformation the
radial equation can be rewritten as a standard wave equation form.
By combining with the real boundary condition (cosine function
form), the harmonic wave appears. Otherwise, for the compactness of
the tortoise coordinate the waves pile up near the horizons
naturally. These effects in the Nariai case in the Brevik's
work\cite{Brevik} are subtle and subclinical  near the cosmological
horizon $r_{c}$.  After rebuilding the scalar field in this paper,
we find that there is no singular peak in waves, and the waves pile
up near the two horizons $r_{e}$ and $r_{c}$, which refreshes
previous works in Refs.\cite{Brevik,Tian}

3 For the widely separated horizons case, the dimensional version of
$\Lambda = 10^{-3}$ reads
\begin{equation}\label{dimensionalversion}
    \Lambda \left(\frac{GM}{c^2}\right) = 10^{-3},
\end{equation}
where $G$ is the gravitational constant, $c$ is the speed of light
and $M$ is the mass of black-hole. Considering a usual Stellar
Black Hole, we assume the mass $M$ equal to ten sun masses (i.e.
$M = 10 M_{\odot}$). It is known that for the Solar system we have
\begin{equation}\label{GMOdot}
    \frac{GM_{\odot}}{c^2} = 1.475km.
\end{equation}
Substituting the mass $M$ and the notation (\ref{GMOdot}) into
Eq.(\ref{dimensionalversion}), we can obtain a dimensional
cosmological constant
\begin{equation}\label{Lambda}
    \Lambda = 4.6 \times 10^{-16} cm^{-2}.
\end{equation}
Using the same method, we can also obtain the dimensional
cosmological constant in the Nariai case.
\begin{equation}\label{LambdaN}
    \Lambda = 2.4 \times 10^{-13} cm^{-2}
\end{equation}
Although the values are much larger than the observed value
$\Lambda_{0}$, it is very necessary to further research them.
After all, the space could be a Schwarzschild one if we take the
observational value $ 10^{-52} m^{-2}$.

\section{acknowledge}Project supported by the National Basic Research Program of China
(Grant No. 2003CB716300) and National Natural Science Foundation of
China (Grant No. 10573003).

\end{document}